\documentclass[sn-mathphys]{sn-jnl}
\theoremstyle{thmstyleone}%

\theoremstyle{thmstyletwo}%

\theoremstyle{thmstylethree}%
\usepackage{hyperref}
\usepackage{threeparttable}
\usepackage[nobiblatex]{xurl}
\usepackage[normalem]{ulem}

\newcommand{\REBOUND}{\texttt{REBOUND}}
\newcommand{\REBOUNDx}{\texttt{REBOUNDx}}
\usepackage[none]{hyphenat}
\allowbreak
\renewenvironment{table}[1][]%
{\tableorg[#1]%
\tablebodyfont%
\renewcommand\footnotetext[2][]{{\removelastskip\vskip3pt%
\let\tablebodyfont\tablefootnotefont%
\hskip0pt\if!##1!\else{\smash{$^{##1}$}}\fi##2\par}}%
}{\endtableorg}

\begin{document}

\title[Risks of ASAT tests in the presence of megaconstellations]{Investigating the risks of debris-generating ASAT tests in the presence of megaconstellations}

\author*[1,2]{\fnm{Sarah} \sur{Thiele}}\email{sarah.thiele@princeton.edu}

\author[1]{\fnm{Aaron C.} \sur{Boley}}\email{acboley@phas.ubc.ca}

\affil[1]{\orgdiv{Department of Physics and Astronomy}, \orgname{University of British Columbia}, \orgaddress{\street{6224 Agricultural Road}, \city{Vancouver}, \postcode{V6T 1Z1}, \state{B.C.}, \country{Canada}}}

\affil[2]{\orgdiv{Department of Astrophysical Sciences}, \orgname{Princeton University}, \orgaddress{\street{4 Ivy Lane}, \city{Princeton}, \postcode{08544}, \state{N.J.}, \country{USA}}}

\abstract{The development of large constellations of satellites (i.e., so-called megaconstellations or satcons) is poised to increase the number of LEO satellites by more than an order of magnitude in the coming decades. 
Such a rapid growth of satellite numbers makes the consequences of major fragmentation events ever more problematic. 
In this study, we investigate the collisional risk to on-orbit infrastructure from kinetic anti-satellite (ASAT) weapon tests, using the 2019 Indian test as a model. 
We find that the probability of one or more collisions occurring over the lifetime of ASAT fragments increases significantly in a satcon environment compared with the orbital environment in 2019. 
For the case of 65,000 satellites in LEO, we find that the chance of one or more satellites being struck by ASAT fragments of size 1 cm or larger is more than 25\% for a single test.
Including sizes down to 3 mm in our models suggests that impacts will occur for any such event. Finally, we apply our methods to examine the November 2021 Russian ASAT test, also finding a significant collision probability over the lifetime of the fragments.
The heavy commercialization of LEO demands a commitment to avoiding debris-generating ASAT tests.}

\keywords{Orbital debris, satellite constellations, N-body integration, Astrodynamics}

\maketitle

\section{Introduction}\label{sec:intro}

	On 27 March 2019, India conducted its first successful anti-satellite (ASAT) test. In this operation, code-named `Mission Shakti', a modified
anti-ballistic missile interceptor was launched by India’s Defence Research and Development Organization (DRDO) to destroy a 740 kg satellite, Microsat-R \cite{langboerk}. 
This event placed India among the three other countries with demonstrated direct ascent ASAT capabilities – United States, Russia and China \cite{chaundry}. 

The mission was implemented with the intent to minimize the amount of long-lasting debris generated by the impact, and thus avoid substantially increasing the risk to crewed space activities, as well as avoid creating a dangerous debris field like the 2007 Chinese ASAT test.
In this vein, the DRDO conducted the test when the target satellite was at a low altitude of 283 km. The satellite was also a relatively small Indian communications satellite, with a surface area of about two square metres \cite{tellis}.

Despite these efforts, more than four months later, there were still 57 tracked and catalogued debris in orbit \cite{grush}. 
Ten of those
57 fragments had apogee altitudes greater than 1000 km, and some as high as 1730 km, spanning the majority of LEO’s altitude range and thus
endangering most LEO spacecraft, including the International Space Station. 
 It is important to recall that
this is only the tracked and catalogued fragments. Smaller debris fragments tend to deorbit faster, so are more likely to remain undetected. However, modelling suggests \cite{jiang} that the number of debris fragments created by Mission Shakti with a size larger than 1 mm may have been of order 10$^5$ -- sizes that could still cause non-negligible damage.

In addition to the risks associated with a direct collision, any resulting fragmentation has the possibility of prompting multiple further fragmentations \cite{kessler}. 
Even if a collisional cascade does not develop, fragmentation events are problematic for the safe operation of  spacecraft in orbit. Furthermore, any fragmentation event adds to the overall number of debris fragments in LEO. The higher the debris density in LEO, the more time missions spend planning and executing avoidance maneuvers. This adds costs in personnel-hours, mission length, and the fuel needed to complete $\Delta$v's \cite{schaub2015, ESA2020, undseth2020}. Most collisions are detected before they occur, so this cost could have a more direct or immediate impact on the space environment than the collisions themselves.

The development of substantial orbital infrastructure, such as `satcons' (i.e., so-called megaconstellations),  makes the negative consequences of  debris-generating ASAT tests much more acute. 
SpaceX already has about 1870 active satellites for its Starlink satcon in orbit\footnote{Based on Celestrak (\url{https://celestrak.com/}), as of 11 February 2022.}, with the potential to grow to 42,000 if the first and second generation configurations are combined \cite{FCCStarlinkInitial, FCCStarlinkVLEO, FCCStarlinkGen1Mod, FCCStarlinkGen2}. Another system, OneWeb, has approximately 390  active satellites in orbit\footnote{\it ibid.}, which could increase to 7,000 by the end of their phase 2 launches \cite{OneWebPhase2}. These substantial changes in the number of spacecraft in LEO create a need for risk evaluation from a cumulative perspective. While the collision probability between a debris fragment and a single satellite may be low, the integrated probability of impact over an entire system could be non-negligible. 

In this study, we explore the collisional risk posed by an ASAT weapon test, similar to that conducted by India in 2019, but with satellites in satcons numbering 65,000 in total. Lastly, in light of the November 15, 2021 Russian ASAT test, we apply our models to investigate  debris cloud lifetimes and subsequent collision probabilities arising from such a test. Because the event took place after the majority of this work was completed, we have included the Russian ASAT test analysis as an appendix.

In regards to terminology, we use  ``megaconstellation" along with ``satcon'' in this study to connect with a wide audience, as the former is currently one of the more recognizable terms. Furthermore, the former emphasizes that these constellations differ from what has been typical for satellite infrastructure, both in physical scale and space environmental impact. However, the reader should be cognizant that the prefix ``mega'' is used colloquially and not technically, and that ``megaconstellation" is losing favour in the technical community. For this reason, the main body of the text uses ``satcon''.

\section{Methods}\label{sec:methods}

We use the astrodynamics code \REBOUND\ \cite{reboundIAS15} to integrate model ASAT-caused debris fragments through satellite density fields, in order to assess the probability that the ASAT test will lead to one or more collisions with a satellite, under different scenarios. 
These density fields are built using public satellite information, as well as proposed satcon configurations. 
Only collisions between the integrated ASAT debris and satellites are considered, i.e., satellite-satellite and impacts from the existing LEO debris field are excluded to focus specifically on the effects of ASAT testing. We note that in a real system,  there could be additional consequences from these primary collisions, such as generating even more debris and raising the risk of yet further collisions (see also Section \ref{sec:intro}).

Details regarding the satellite fields are given below. 
In brief, a satellite distribution is selected and the satellites in that distribution are time-averaged along their orbits. 
LEO is divided into azimuthally symmetric grid cells, according to altitude and co-latitude. 
Then, the altitude and co-latitude positions of the satellites along their orbits are used to assign weighted densities to the grid cells, such that a satellite's contribution to a grid cell depends on the time the satellite spends passing through the cell's volume.

Three satellite distributions are used to create density fields for this study: (1) The satellites in LEO as of August 2019\footnote{Data obtained from USSPACECOM (\url{https://www.space-track.org}).}, (2) the first 12,000 of SpaceX's Starlink constellation, which will be launched over the next five years, and (3) an environment of 65,000 satellites which include the constellations Starlink, OneWeb, Amazon/Kuiper, and GW/StarNet. 

\subsection{\REBOUND\ and \REBOUNDx}\label{sec:rebound}
\REBOUND\ is an open-source, flexible N-body integrator that can be used for a wide range of astrophysical and astrodynamical problems. 
We use the Wisdom-Holman integrator \textsc{WHFast}  \cite{reboundWHFAST} for all simulations presented here, which is well-suited for cases in which orbital perturbations are small. 
As a consistency check,  \REBOUND's  \textsc{IAS15} integrator \cite{reboundIAS15} was also used in preliminary test simulations, which gave statistically indistinguishable results from the \textsc{WHFast} runs.

Gas drag is implemented by including a force term
\begin{equation}
    {\bf F}_d=-\rho \frac{C_{\rm D}}{2} \left({\rm A/M}\right)v~{\bf v}
\end{equation} 
on all fragments. 
The velocity of a debris particle ${\bf v}$ (and its magnitude $v$) relative to the geocentre  is also taken to be the particle's speed through the atmosphere for the purpose of determining drag forces.

The drag coefficient C$_{\rm{D}}$ is held constant at 2.2. 
The gas density $\rho$ is a function of altitude only and is determined by linearly interpolating, in log space, between the 2012 tabulated values of the COSPAR International Reference Atmosphere (CIRA)  \cite{CIRA2012}. For objects at altitudes greater than those included in CIRA-2012, the density is linearly extrapolated, in log space, from the two highest entries in the density table. Our \REBOUND\ gas drag algorithm accounts for thermospheric density variation based on the solar cycle by interpolating between the CIRA-2012 low and high solar activity profiles using $\sin( 2\pi t / (22~\rm yr) +\phi_0)^2$, but does not take into account diurnal variation. Simulations can begin at an arbitrary phase $\phi_0$ of the solar cycle. The Indian ASAT test took place only nine months before solar minimum ($\phi\simeq0.93\pi$), which is included in the modelling. The area-to-mass ratio A/M is discussed further below.

 Collision probability is calculated as a Poisson probability. During each integration timestep $dt$, the instantaneous collision probability is determined by 
 \begin{equation}
     p= n \sigma v_{\rm{rel}} dt\label{eq:instant_colprob}
 \end{equation}  
 where $n$ is the local satellite number density,  $\sigma=10~\rm m^2$ is the assumed satellite cross-section (note that the results are scalable by this value), and $v_{\rm{rel}}$ is the typical relative speed between an ASAT fragment and the satellites in the given grid cell.
 Because the probability of any one fragment having a collision is small, the cumulative collision probability for the $i^{th}$ particle, $\lambda_i$, is the sum of an individual particle's instantaneous collision probabilities up until the simulation ends or the particle de-orbits. Finally, assuming that each particle is independent, the total probability of having one or more collisions as a result of the ASAT test under Poisson statistics is 
 \begin{equation}
     P_{coll}=1-\exp\left(-\Lambda\right)\label{eq:colprob}
 \end{equation}
 for $N$ debris fragments and $\Lambda = \sum_i^N\lambda_i$.
 
As we use a density field for the satellites, we need a statistical approximation for $v_{\rm{rel}}$. 
If all orbits are approximately circular and in the same shell, then the average relative velocity is $v_{\rm{rel}}=\frac{4}{\pi} v_{\rm{c}}$, where $v_{\rm{c}}$ is the local circular orbital speed\footnote{The value 4/$\pi$ assumes that the relative velocities are isotropically distributed. If we instead assume that the orbits are all isotropically distributed, then the averaging is slightly different, and a value of 4/3 should be used. Because these values are close, we maintain the 4/$\pi$ value.}. Because the fragments can have non-trivial eccentricities, we modify the relative speed to be
\begin{equation}\label{eq:relvel}
    v_{\rm{rel}}\approx \left( \frac{16}{\pi^2} v_\theta^2 + v_r^2\right)^{1/2},
\end{equation}
for azimuthal and radial velocity components $v_\theta$ and $v_r$, respectively, of the individual debris fragment. 
At low eccentricities, Equation \ref{eq:relvel} is the same as that for circular orbits. 
In tests, the relation gives values that are within about 10\% of those determined from Monte Carlo averaging, even for high eccentricity.

\subsection{Satellite Density Field}\label{sec:satdis}
In order to simplify the collision probability calculations, we use azimuthally symmetric\footnote{In this study we refer to ``azimuth" in the customary sense pertaining to spherical coordinates. That is, we use the coordinates ($R,\theta,\phi$) as the radial, polar, and azimuthal coordinates respectively to describe positional vectors, where the azimuthal angle is measured from an arbitrary reference point.} spherical grid cells to represent an averaged satellite density field. 
This avoids the need for directly calculating a collision probability between debris fragments and each individual satellite, an expensive computational process. The density field is thus held constant during the simulation.

We assign each cell $i$ an altitude $h_i = R_i - R_\oplus$ and a co-latitude $\theta_i$ value, with respective widths of $\Delta R = 1$ km and $\Delta \theta = 1^\circ$. 
The inner boundary of the grid has an altitude $h_{\rm{min}}=300$ km. 
For the satcon distributions, the upper grid boundary is $h_{\rm{max}} = 1,500$ km. 
However, the comparison simulation that uses the actual 2019 satellite distribution requires a higher upper boundary, with $h_{\rm{max}} = 2,000$ km.
Note that these boundaries are only for defining the satellite density fields.  

Because all cells are azimuthally averaged, a single cell has a volume of 
\begin{equation}
    V_i = 2\pi R_i^2 \Delta R[\cos{(\theta_i-\Delta \theta/2)}-\cos{(\theta_i+\Delta \theta/2)}]
\end{equation}

The orbital parameters for satcons are taken from FCC filings and are listed in Table \ref{tab:satparams}. 
Each row lists satellites with like altitudes and inclinations. 
N$_{\rm{P}}$ is the number of unique orbital planes occupied by that particular set of satellites. N$_{\rm{SPP}}$ is the number of satellites per plane. If this value is 1, each satellite occupies its own orbital plane and the satellites are distributed evenly along their orbits. 

\begin{table}
\begin{center}
    \begin{threeparttable}
    \centering
	\caption{Parameters for future LEO satcons, obtained from FCC filings. The number of unique orbital planes is N$_{\rm{P}}$ with number of satellites per plane N$_{\rm{SPP}}$. All satellites in a given row have the same orbital inclination and altitude. }
	\label{tab:satparams}
	\begin{tabular}{l*4c}
	\\
		\hline
		\hline
		\\
		Constellation & N$_{\rm{P}}$ & N$_{\rm{SPP}}$ & Inc. & Alt. \\
		& & &($^o$)&(km)
		\\
		\hline
		\hline
		\\
		& 2547 & 1 & 53 & 345.6 \\
		& 2478 & 1 & 48 & 340.8 \\
		& 2493 & 1 & 48 & 340.8 \\
		Starlink Gen1\tnote{a}& 72 & 22 & 53 & 550 \\
		& 72 & 22 & 53.2 & 540 \\
	    & 36 & 20 & 70 & 570 \\
        & 6 & 58 & 97.6 & 560\\
        & 4 & 43 & 97.6 & 560.1\\
		 \\
		\hline
		\\
		 & 7178 & 1 & 30 & 328 \\
		 & 7178 & 1 & 40 & 334 \\
		 & 7178 & 1 & 53 & 345 \\
		 Starlink Gen2\tnote{b} & 40 & 50 & 96.9 & 360 \\
		 & 1998 & 1 & 75 & 373 \\
		 & 4000 & 1 & 53 & 499 \\
		 & 12 & 12 & 148 & 604 \\
        \\
        \hline
        \\
        & 18 & 40 & 87.9 & 1200\\
        OneWeb\tnote{c}& 36 & 49 & 87.9 & 1200\\
        & 32 & 72 & 40 & 1200\\
        & 32 & 72 & 55 & 1200\\
        \\
        \hline
        \\
        & 16 & 30 & 85 & 590\\
        & 40 & 50 & 50 & 600\\
        & 60 & 60 & 55 & 508\\
        GW/StarNet\tnote{d} & 48 & 36 & 30 & 1145\\
        & 48 & 36 & 40 & 1145\\
        & 48 & 36 & 50 & 1145\\
        & 48 & 36 & 60 & 1145\\
        \\
        \hline
        \\
        & 34 & 34 & 51.9 & 630 \\
        Amazon/Kuiper\tnote{e} & 36 & 36 & 42 & 610 \\
        & 28 & 28 & 33 & 590 \\
        \\
		\hline
        \hline
	\end{tabular}
    \begin{tablenotes}
        \item[a] Approximately 12,000 satellites \cite{FCCStarlinkInitial, FCCStarlinkVLEO, FCCStarlinkGen1Mod}.
        \item[b] Assumes Gen 1 and 2 operating together for a total of approximately 42,000 satellites \cite{FCCStarlinkGen2}.
        \item[c] Includes phases 1 and 2 for approximately 7,000 satellites \cite{OneWebPhase2}.
        \item[d] 12,992 satellites, based on multiple ITU filings \cite{StarNet}.
        \item[e] 3,236 satellites \cite{Kuiper}.
    \end{tablenotes}
    \end{threeparttable}
    \end{center}
\end{table}

For creating the distributions, two different methods are used. In both methods, all satellites in the distribution are added at the same time, mimicking a LEO environment in which the entire chosen population is fully established in orbit. For satcons, we distribute the nodes evenly between 0 and $2\pi$. The satellites are then placed randomly along their orbits in their respective plane. Due to this randomization, some clumping occurs in the corresponding satellite density fields if only instantaneous positions are used. As a result, the satellites are integrated in \REBOUND\ and the density field is created by averaging multiple snapshots. For both the 65,000 and 12,000 satellite case, we found that 100 snapshots over an orbital period at 600 km altitude is sufficient to yield a smooth distribution.

A different method is used for defining the 2019 satellite density field (i.e., only the satellites that were in orbit at that time). In this case, the orbital parameters for a 2019  satellite distribution are obtained from the USSPACECOM catalogue for August 2019. 
We only include satellites with perigees below an altitude of 2000 km. This smaller number of satellites ($\approx 3000$ for 2019) again requires time-averaging for determining a smooth satellite density distribution. In this case, we use each satellite's osculating Keplerian orbits and weight its contribution to each cell by the amount of time it spends in that cell over an orbit. 

\begin{figure}
    \centering
    \includegraphics[scale=0.0475]{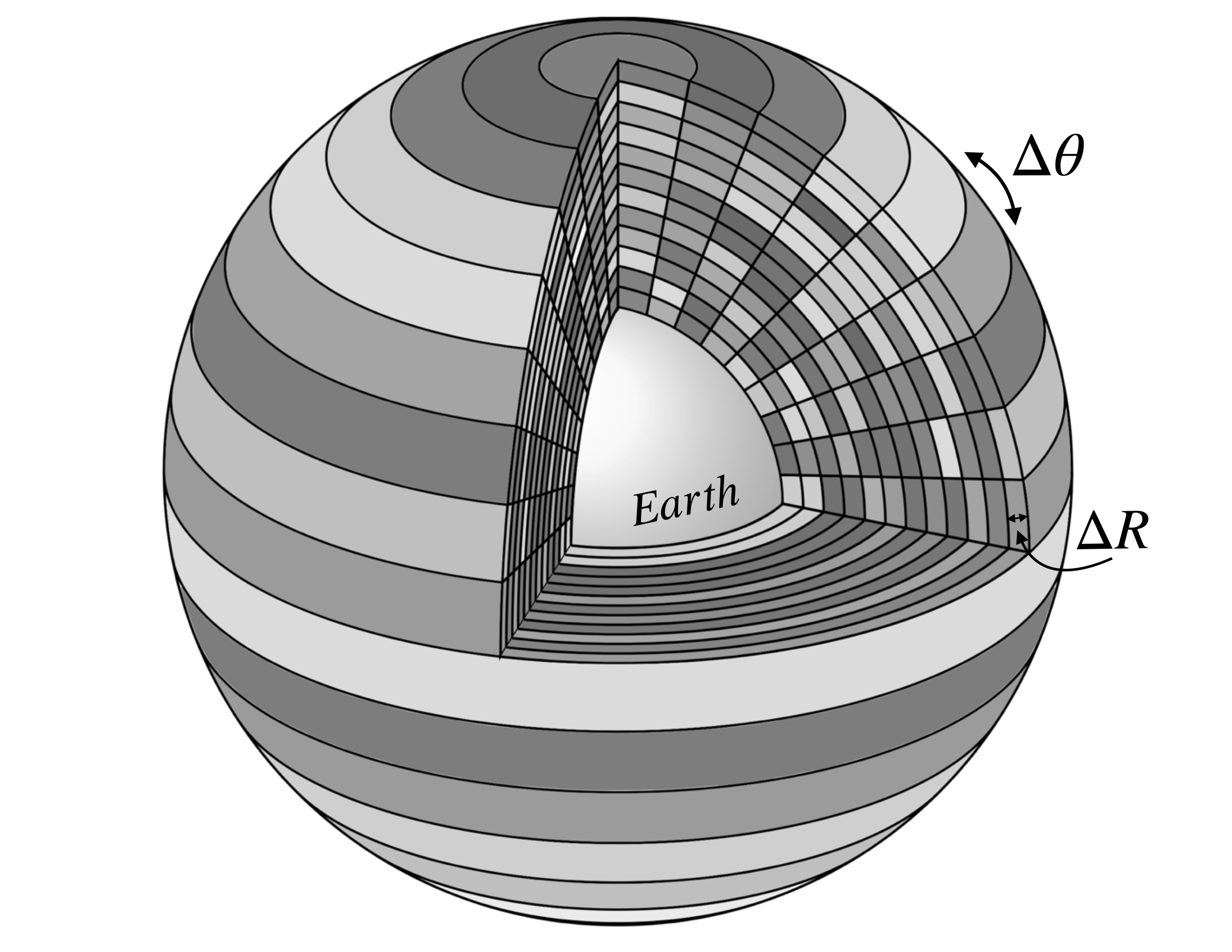}    \caption{An illustration of the grid used for the satellite density field. The altitude and co-latitude cell widths $\Delta R$ and $\Delta\theta$, respectively, are shown.  Grey shading is used to represent an arbitrary orbit-averaged satellite density field, again for illustrative purposes only. Because the cells are azimuthally symmetric, the density field appears as a distribution of rings, with each ring having its own density.}
    \label{fig:densitygrid}
\end{figure}

An illustrative representation of the grid and satellite density field is shown in Figure \ref{fig:densitygrid}. The Earth is at the centre, with grid cells set above Earth's surface and characterized by a radial width $\Delta R$ and co-latitude width $\Delta \theta$. The shading is used to represent a toy distribution of satellite densities for illustrative purposes only. The actual satellite density for each cell $i$ is calculated as described above.

\subsection{Simulating a Kinetic ASAT Test}
We investigate two methods of simulating debris distributions arising from ASAT testing.  The first uses the NASA Standard Breakup Model (NSBM), in which the distributions for fragment velocity kicks and area-to-mass ratios A/M are generated in a probabilistic manner.
The second is a simplified approach using a Rayleigh distribution of fragment velocity kicks and a fixed A/M ratio. This is intended to aid in the interpretation of the results and to provide a basic model for easy reproducibility.  
We discuss these methods in turn. 

\subsubsection{NASA Standard Breakup Model}\label{sec:SBM}
The NSBM is implemented following \cite{NSBM}, which yields probability distributions for parameters that describe debris formed from an explosion or collision. 
The distributions that we use correspond to the parameterizations for a collision. 
The choice is intended to better represent the energetics of the Indian 2019 ASAT test, which used a kinetic kill vehicle rather than an explosive warhead. Distribution parameters can be found in Appendix B of \cite{NSBM}.

In the NSBM, the number of fragments with lengths greater than a given characteristic length $L_{\rm{c}}$ is given by:
\begin{align}
    N(> L_{\rm{c}}) &= 0.1 M_{\rm{e}}^{0.75}L_{\rm{c}}^{-1.71} \\ 
    \text{or } \Delta N &= N(> L_0) - N(> L_1) \label{eq:dN}
\end{align}
where $M_{\rm{e}}$ is the mass of ejected fragments in kilograms, $L_{\rm{c}}$ is the fragment's characteristic length, in metres, and $L_0$ and $L_1$ are the lower and upper bounds on a given $L_{\rm{c}}$ bin, respectively. 

The fragmenting mass $M_{\rm{e}}$ caused by the collision is dependent on the kill energy per target mass, which is
\begin{equation}\label{eq:KEm}
    \epsilon = \frac{1}{2}\frac{m_i}{m_t}v_i^2 = \frac{E_{\rm{kill}}}{m_t}\rm ,
\end{equation}
where $\rm m_i$ and  $\rm m_t$ are the interceptor and target mass, respectively, $v_{\rm{i}}$ is the relative speed between the interceptor and target, and $E_{\rm{kill}}$ is the chosen kill energy. The associated relative impact speed is thus $v_{\rm{i}}$ = $\sqrt{2E_{\rm{kill}}/m_i}$.

$\text{If }\epsilon < 40$ J/g, it is deemed a `noncatastrophic' collision, so the target is only cratered, and the ejected mass of fragments is a scaling of the kill energy $M_{\rm{e}} = m_i\big{(}\frac{v_i}{\rm km/s}\big{)}^2$. $\text{If }\epsilon \geq 40$ J/g, it is a catastrophic collision and the fragment mass that is ejected is the total mass of the interceptor and target $M_{\rm{e}} = m_i + m_t$.

We use an effective energy of 130 MJ for India's ASAT test (based, in part, on providing correspondence between Gabbard plots -- see below), yielding a catastrophic collision with a fragment mass of $M_{\rm{e}} = 750$ kg. 
This mass includes 740 kg for the Microsat-R satellite and 10 kg (estimated) for the kill vehicle. We use 100 logarithmically spaced $L_{\rm{c}}$ bins between 3 mm and 1 m, and with Equation \ref{eq:dN} obtain a number distribution as a function of size for the generated debris fragments.

The logarithm of A/M ratios are sampled from unique probability distribution functions for each size, which are linear combinations of normal distributions detailed in the NSBM. The weights, means, and standard deviations for those distributions are functions of $\log_{10}(L_{\rm{c}})$ and depend on the size regime of the fragment. For large fragments with $L_{\rm{c}} > 11$ cm, the PDF is bimodal, small fragments with $L_{\rm{c}} < 8$ cm are normally distributed, and sizes between these two regimes incorporates a transition function that is a linear combination of the two previous PDFs. These PDFs are sampled to obtain a distribution of A/M ratios. The area and mass for the fragment is obtained using its A/M ratio and Equations (42a,b) of \cite{NSBM}. As discussed in \cite{NSBM}, the NSBM does not intrinsically conserve physical quantities like mass and  energy. In order to enforce conservation of mass, the initial number distribution $\Delta N(L_{\rm{c}})$ is normalized using the ratio of the target payload mass to the total fragment mass $m_t/m_{\rm{frag,tot}}$, the A/M and mass distributions are re-sampled, and this process is run iteratively until $m_{\rm{frag,tot}} \leq M_{\rm{e}}$, resulting in a total number of fragments $N_{\rm{tot}}$ and an updated  mass of fragments $M_{\rm{e}}^*=m_{\rm{frag,tot}}=\sum_{i=1}^{N_{\rm{tot}}}m_{\rm{frag,}\textit{i}}$.

This process yields a number of fragments that is $\mathcal{O}(10^5)$ for $L_c$ $\in [0.003, 1.0]$ (unit of metres implied), an unrealistic amount for long-term integration. We thus perform integrations of a sampling of 1,000 fragments. A few simulations were executed with 5,000 fragments as well to confirm that 1,000 gives a representative parameter set. These fragments are sampled from the distribution of A/M ratios that results from the above process. Using these A/M ratios, the NSBM gives a unique normal probability distribution of $\log_{10}(\Delta v)$ for each fragment $i$ with a mean $\mu = 0.9\log_{10}(A/M)_i+2.9$ and standard deviation $\sigma = 0.4$ (using mks units). A value is sampled for each fragment from its unique PDF to obtain a distribution of $\Delta v$'s for the fragments. These magnitudes are combined with a unit vector generated randomly from a uniform distribution between -1 and 1 for each component, resulting in the final velocity kicks. 
In order to perform one simulation in which the 1000-fragment sample is close to a full NSBM number distribution, we also simulate a debris cloud using 100 characteristic length bins between 10 cm and 1 m. The size of 10 cm also represents the approximate lower limit to the size of most trackable debris in LEO (although with technological advances in radar detection, this bound is improving towards fragment sizes of $\mathcal{O}(1 ~\rm cm)$). 

Our generated values for A/M ratio and $\Delta v$'s are then integrated in \REBOUND\ and their collision probabilities recorded, as outlined in Section \ref{sec:integrate}.

\subsubsection{Rayleigh Distribution}\label{sec:rayleigh}
In this simplified method, we assume that the magnitudes of the fragment velocity kicks approximately follow a Rayleigh Distribution. 
We further fix the A/M ratio and the total number of fragments in the simulation, sidestepping any need to assign characteristic lengths or specific fragment masses. 
 To compare this method with the NSBM, we similarly perform simulations with 1,000 fragments, and later scale the resulting collision probabilities (see Section \ref{sec:integrate}) using the fragment numbers found by the mass-normalized number distribution using Equation \ref{eq:dN} with $L_{\rm{c}} \in [0.003, 1.0]$ and $L_{\rm{c}} \in [0.1, 1.0]$.

We set the mode of the Rayleigh distribution to $\sigma = 250$ m/s and draw from the distribution for the magnitude of the velocity kicks, which produces a reasonable correspondence to the actual Gabbard plot of the Indian ASAT test.  A randomly oriented unit vector is assigned to each fragment, as in the NSBM. The fragments are added to a \REBOUND\ simulation and integrated using the same process as described in the following section.

In this method, simulations are run for both A/M = 0.05 m$^2$kg$^{-1}$ and A/M = 0.04 m$^2$kg$^{-1}$ for all particles, unless stated otherwise. We  refer to each simulation set as model A/M0.05 and A/M0.04, respectively.

\subsection{Debris Integration}\label{sec:integrate}

Table \ref{tab:microsatR} lists the orbital parameters that we use for Microsat-R, the satellite destroyed in the Indian ASAT test. For each \REBOUND\ simulation, the fragments are initialized with these orbital parameters, mimicking the instantaneous moment just prior to the fragmentation event. 
Each fragment is then perturbed following ${\bf v} = {\bf v}_{\rm{0}} + \Delta {\bf v}$, where ${\bf v}_{\rm{0}}$ is Microsat-R's approximate velocity just before the collision and $\Delta {\bf v}$ is a fragment's respective velocity kick. This further represents $t=0$ of the simulation; as the satellite density field is static with time, no specific epoch is required. The only time-dependent parameter is the solar phase $\phi$, which is discussed further in Section \ref{sec:rebound}.

\begin{table}[h]
    \centering
	\caption{Orbital parameters of Microsat-R just prior to the Indian ASAT test \cite{greatgame}.}
	\label{tab:microsatR}
	\begin{tabular}{l*2c}
	\\

		\hline
		\hline
		\\
		Altitude of ASAT test & 283 km \\
        Perigee & 267.4 km  \\
        Apogee & 288.7 km  \\
        Inclination & 96.6°  \\
        Period & 89.9 min  \\
        Semi-major axis & 6649 km  \\
        \\
		\hline
        \hline
	\end{tabular}
\end{table}

In addition to gas drag, we include the Earth's $J_2$ and $J_4$ components. When the simulations are initialized, the $\beta$ coefficients for gas drag are calculated from the A/M ratio distribution for all fragments using $\beta = C_D  $A/M. 
For the Rayleigh distribution, $\beta=0.11$ and $\beta=0.088$ for an A/M0.04 and A/M0.05, respectively for all fragments. At the start of a simulation, a desired satellite density field (see Section \ref{sec:satdis}) is selected and read into \REBOUND$.$ The fragments are then integrated until either all the fragments have de-orbited or the simulation reaches three years of integration, whichever happens first. After every 30 minutes in the simulation, \REBOUND\ removes particles that have decayed below a threshold altitude, which we set to 200 km. 

When a fragment de-orbits, its cumulative collision probability $\lambda$ is recorded for evaluating the total collision probability $P_{coll}$. Likewise, after three years of integration time, any remaining fragments also have their $\lambda$'s included in the calculation of $P_{coll}$. 

For a particular satellite density field, the collision probability scales approximately linearly with the number of fragments, provided that the number of fragments adequately samples the full fragment distribution.
This is the case with the simulations run here, which was verified by running identically seeded simulations with $N_{\rm{frags}}=(1,2,3,4,5)\times 10^3$ for both the NSBM and Rayleigh methods. For any simulation, we can thus scale the cumulative collision probability of a simulated sub-sample of fragments by the total number of fragments that would result from an entire NSBM number distribution. This scaling will be discussed in further depth later in Section \ref{sec:results}.

\section{Results: Modeling an India ASAT-like Test}\label{sec:results}
\subsection{ASAT Debris Field Initial Orbital Distributions}\label{subsec:results_initdis}
Using the NSBM for ${\rm 3~mm} < L_c < 1~{\rm m}$ yields more than 300,000 debris fragments, with the majority of these below a few centimetres in size. 
This is consistent with other studies of Mission Shakti \cite[e.g.,][]{jiang}, which find  debris numbers of order 10$^5$ for sizes greater than 1 mm. 
The NSBM gives a re-normalized fragment number of $N_{\rm{frag}}=1168$ for $L_{\rm{c,min}}=10$ cm. 
While this value is significantly smaller than including sizes down to 3 mm, it is noteworthy that the consequences of a $10+$ cm-sized fragment colliding with a spacecraft will be much more consequential than that of a millimetre-sized fragment. Thus while the lower fragment number for $L_{\rm{c,min}}=10$ cm may yield a lower collision probability, the result of the collision itself is expected to be of higher significance.

\begin{figure}
    \centering
    \includegraphics[scale=0.077, trim={10cm 0cm 10cm 0cm}]{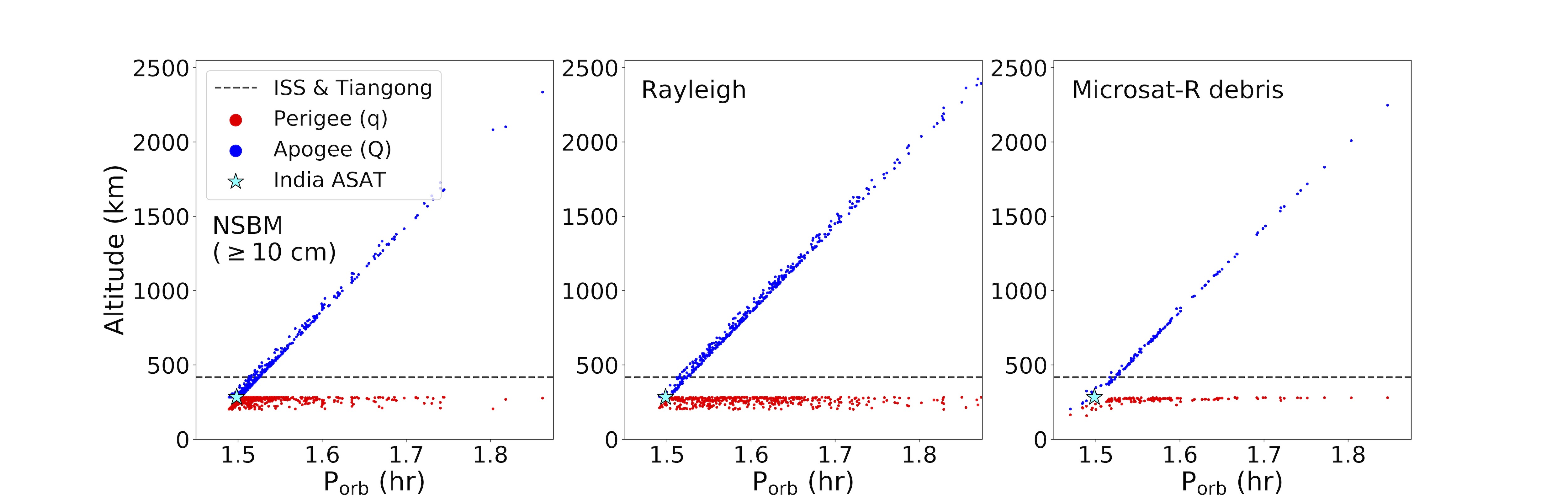}
    \caption{Gabbard plots for low-altitude kinetic ASAT tests. From left to right, the plots show the initial state of the NSBM simulation with $L_{\rm{c,min}}=10$ cm, the Rayleigh $\Delta v$ distribution, and the actual fragment data from the 2019 Indian ASAT test, based on the first available TLE for each fragment (data: USSPACECOM). Fragment apogees are shown in blue, and the perigees in red. The altitude of the International Space Station and China's Tiangong space station is shown as a dashed line (note: this is intended to delineate altitude, only). The orbital period and altitude of Microsat-R satellite at the time of the Indian ASAT test are shown as the blue star. }
    \label{fig:gabbardgrid}
\end{figure}

Figure \ref{fig:gabbardgrid} shows Gabbard plots for three fragmentation models. The left panel is for a NSBM simulation with $L_{\rm{c,min}}=10$ cm, while the middle panel is for the Rayleigh method, both with 1,000 fragments. 
Both panels are for the debris immediately following the modelled fragmentation event. 
Only fragments with apogees below 2,500 km are shown for comparison with the right panel, which shows data from the Indian ASAT test (i.e., Microsat-R debris)\footnote{Data are the first available TLE for each particle, as provided by USSPACECOM.}. Many of these fragments first appear in the satellite/debris catalogue well after the ASAT was conducted and therefore already include substantial orbital evolution due to gas drag.  Thus, some differences in comparison with the other two panels are expected. As their orbits evolve, the fragments migrate toward the bottom left corner of the plot, as can be seen in the smattering of points in the bottom left of the Microsat-R debris panel. The NSBM finds a mass-renormalized $N_{\rm{frag}}=1168$ for $L_{\rm{c,min}}=10$ cm, of which there are 647 that do not immediately de-orbit and have an apogee $Q$ below 2500 km. The 1000-fragment simulation used for Figure \ref{fig:gabbardgrid} similarly yields 539 fragments with $Q<2500~\rm{km}$ and $L_{\rm{c,min}}=10$ cm. From the actual Indian ASAT test, there were only 129 fragments tracked in USSPACECOM. The discrepancy may be due to fragments of trackable size being missed (including deorbiting before being catalogued), the breakup being different from the NSBM, or some combination of both.

\begin{figure}
    \centering
    \includegraphics[scale=0.0475]{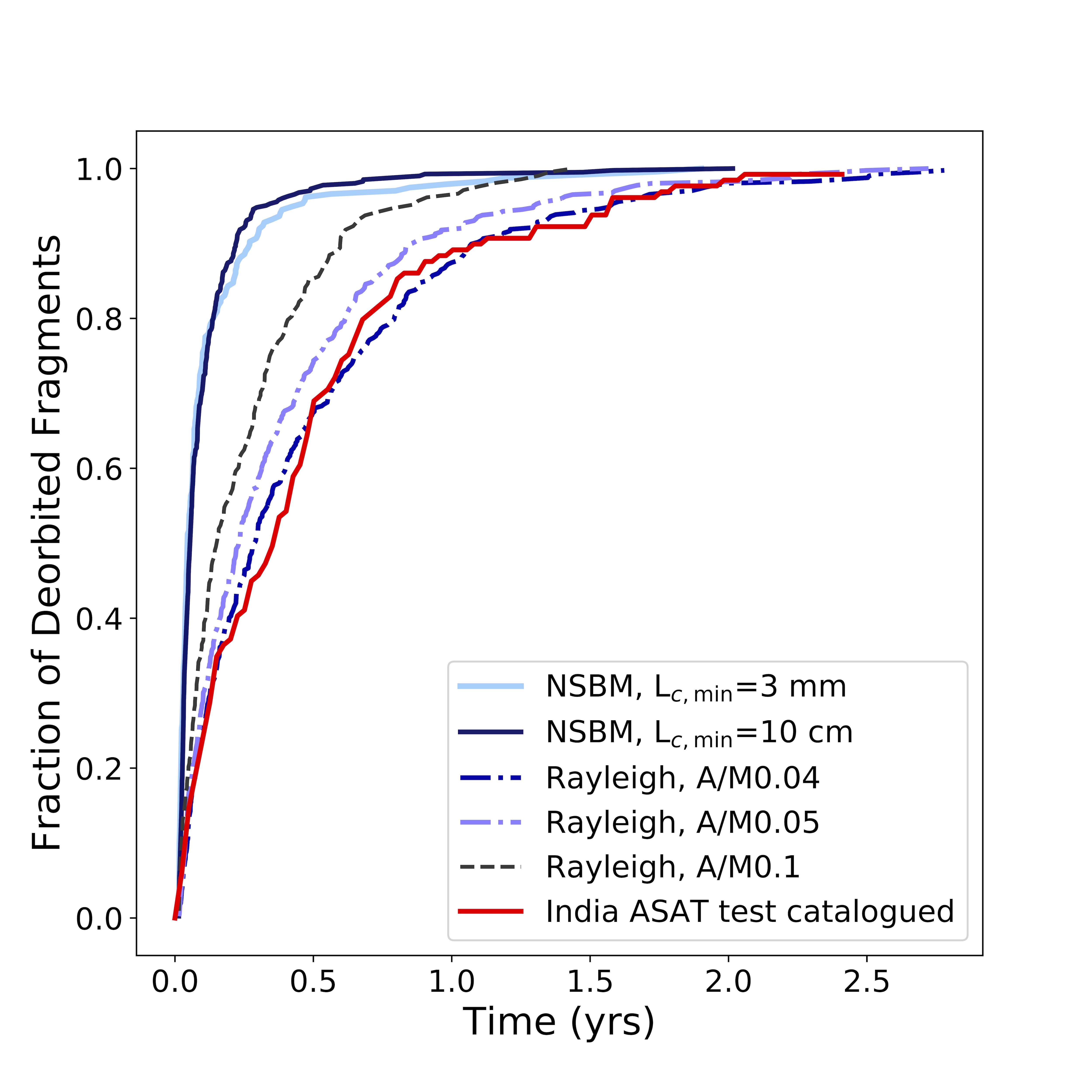}
    \caption{The fraction of de-orbited fragments as a function of time. Simulations are run using 1,000 fragments. Moreover, only fragments with lifetimes greater than five days are shown. The NSBM with $L_{\rm{c,min}}$ = 3 mm and 10 cm is represented by solid light and dark blue lines, respectively. Three Rayleigh-method simulations are also shown: our standard 1,000 fragment simulations with A/M = 0.05 m$^2$kg$^{-1}$ (dash-dotted royal blue curve) and A/M = 0.04 m$^2$kg$^{-1}$ (purple dash-dotted curve), as well as a lower-resolution model with 500 fragments and A/M = 0.1 m$^2$kg$^{-1}$ (black dashed line), which was done to demonstrate the sensitivity of the de-orbit times to the A/M ratio. All fragments de-orbit by the maximum integration time of three years. The re-entry data for Indian ASAT test fragments, as given in the USSPACECOM catalogue, are shown as a red solid line.}
    \label{fig:deorbitfrac}
\end{figure}

\subsection{Debris De-orbit Timescales and $\Delta v$ Distributions}\label{subsec:results_deorbits}
Figure \ref{fig:deorbitfrac} shows the fraction of fragments that have de-orbited as a function of time. 
Only fragments that de-orbit after the first five days are included to focus on long-lasting debris. Five different simulation results are shown: the NSBM simulations with minimum characteristic lengths of 3 mm and 10 cm, and three simulations using Rayleigh distributions, one with a constant ${\rm A/M} = 0.05 \rm ~m^2~kg^{-1}$ and another with ${\rm A/M} = 0.04 ~\rm m^2~kg^{-1}$ (models A/M0.05 and A/M0.04, respectively). Another simulation with ${\rm A/M} = 0.1 ~\rm m^2~kg^{-1}$ (henceforth model A/M0.1) is run with a lower resolution (500 fragments) in order to demonstrate the sensitivity of the fixed A/M ratio in the Rayleigh simulations.  These are all compared with the actual de-orbit times for the tracked Microsat-R debris. 

The fragments in all simulations completely de-orbit before the end of three years, except for a single fragment in model A/M0.4. The Rayleigh simulations, especially A/M0.04, show a much shallower slope in the de-orbit fraction in comparison to A/M0.1 and the NSBM. This result visualizes the sensitivity of the simulations to the A/M ratio distribution, all other things being equal. For reference, the peaks of the A/M ratio distributions in the NSBM models range from 0.1 to 0.5 m$^2$ kg$^{-1}$, depending on $L_{\rm{c,min}}$. 
Fragments that were ejected into hyperbolic orbits are not included in the de-orbit fraction, but are worth noting. The NSBM $\Delta v$ distribution has a high velocity tail, especially for simulations with the lower $L_{\rm{c,min}}$ value of 3 mm. While relatively rare, the hyperbolic trajectories tended to occur in the NSBM for sizes at or below 1 cm. 
For example, 20 out of 1,000 fragments were ejected from Earth orbit in the $L_{\rm{c,min}}$ = 3 mm NSBM simulation, all smaller than 1.5 cm.
The Rayleigh distributions did not show this behaviour.

In comparison with the Indian ASAT test at face value, both NSBM simulations and the Rayleigh A/M0.1 models have shorter de-orbit timescales than what is catalogued for the Microsat-R debris.
It must be kept in mind that the data are biased, in part, by the latency in identifying the debris from Microsat-R. 
For example, only about half of the fragments that were eventually catalogued by USSPACECOM were identified within approximately two weeks following the event. 
Thus, substantial evolution of the fragment ensemble, including de-orbits, likely occurred before their identification (or inclusion in the catalogue).
The catalogued fragments are also incomplete. 
Nonetheless, these biases do not necessarily account for Microsat-R debris having an overall longer de-orbit timescale than what we see in our simulations.
Initially, it might seem that incorporating a larger effective kill energy into the NSBM would lengthen de-orbit timescales by ejecting fragments onto higher apogees. However, the kill energy in the NSBM mainly serves to differentiate between a catastrophic and non-catastrophic collision. While the latter implicitly incorporates the kill energy through the determination of the ejected mass, once the threshold energy for a catastrophic collision is reached, the kill energy does not further alter the distribution. 
In the case of the Indian ASAT test, the NSBM does not adequately model the properties of the long-term Microsat-R debris. Follow-up studies might need to incorporate kill energy in other ways, such as imposing equipartition of energy into the fragment sampling on top of the mass conservation we employ in this work.

We stress that the simulations are not intended to reproduce exactly the Indian ASAT test; rather the comparison is intended to show the plausibility and relevance of the models we explore. 
The faster de-orbit times seen in most of the simulations, again at face value, suggest that the collision probabilities derived from them underestimate the collision probability that was posed by the Indian ASAT test. 

The Rayleigh models A/M0.04 and A/M0.05 produce a more promising fit to the catalogued debris. The slope of each model follows a similar trend to the observed data, although exhibit slightly longer de-orbit timescales, by a few fragments-worth, up to $\sim 3$ years. This may suggest that a Rayleigh method with these A/M ratios is able to capture some of the observational biases in the catalogued debris, or is simply a better overall model for this particular event than the NSBM.

\begin{figure}
    \centering
    \includegraphics[scale=0.07, trim={4cm 0 4cm 0}]{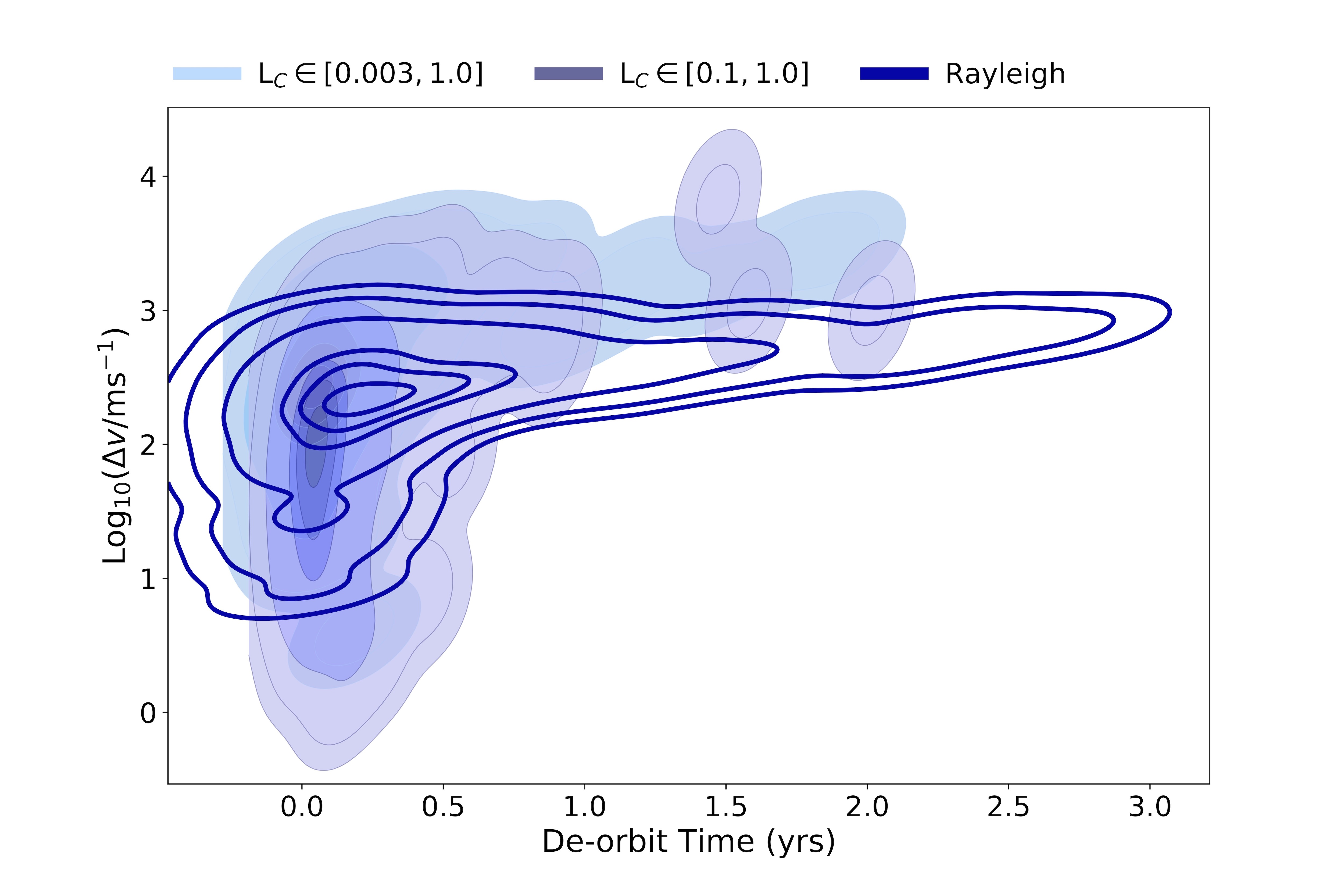}
    \caption{The fragments' $\Delta v$ velocity kicks shown against their de-orbit times. Contours represent 0.1\%, 1\%, 10\%, 50\%, 70\%, 90\%, and 100\% of the fragments for each fragmentation model. The NSBM simulations for $L_{\rm{c,min}}$ = 3 mm and 10 cm are shown in the light and medium blue shading, respectively. 
    Each simulation includes a sample of 1,000 fragments. The Rayleigh distribution contours are shown in dark blue (without fill) and uses 1000 fragments. Only A/M0.04 is shown here, but we find when A/M0.05 is overlaid that the two distributions are quite similar, with A/M0.04 having de-orbit times about 0.1 years longer.
    Only fragments that de-orbit past five days are shown. The majority of fragments deorbit in under six months, especially those with low velocity kicks whose orbits remain at low altitudes and eccentricities. The Rayleigh model produces longer de-orbit timescales on average since the NSBM has A/M ratios which peak at higher values than 0.05 m$^2$kg$^{-1}$.}
    \label{fig:v_t_kde}
\end{figure}

The relation between velocity kicks and de-orbit timescales is shown in Figure \ref{fig:v_t_kde}. The contours for each simulation represent seven levels between 0.1\% and 100\% of $\log_{10}(\Delta v)$. Only fragments with $t_{\rm{deorbit}}>5$ days are included in this plot. As expected, most fragments de-orbit within the first few months following the ASAT test. This effect is exaggerated for the NSBM, with fragments at the extremes of the velocity kicks de-orbiting quickly and adding to the overall short-term de-orbit fraction like that seen in Figure \ref{fig:deorbitfrac}. The $\Delta v$ distributions cluster between $10^2-10^3$ m/s for the NSBM simulation with $L_{\rm{c}}\in [0.003,1.0]$. 
The mode of the Rayleigh distribution overlaps the NSBM distribution, although the NSBM extends to higher $\Delta v$ values, which is consistent with the differences in the models. The NSBM simulation with $L_{\rm{c}}\in [0.1,1.0]$ also shows a concentration of $\Delta v$ values around 50 to 150 m/s. Again, in the NSBM, larger fragments tend to have lower A/M ratios, and hence lower $\Delta v$ values on average.

\begin{table}[h]
    \centering
	\caption{Collision Probabilities: India ASAT-like test}
	\label{tab:results}
	\begin{tabular}{l*6c}
	\\
		\hline
		\hline
		\\
		\multicolumn{1}{l}{\textbf{Model}} & \multicolumn{2}{c}{~\textbf{2019 sats}} &  \multicolumn{2}{c}{~\textbf{12,000 sats}} & 
 \multicolumn{2}{c}{~\textbf{65,000 sats}} \\
		\\
		& $P_{coll}$ & $P_{coll}$ &  $P_{coll}$ & $P_{coll}$ &  $P_{coll}$ & $P_{coll}$ \\
		&  \multirow{2}{1cm}{$N_{\rm{sample}}$ = 1000} & Scaled  &   \multirow{2}{1cm}{$N_{\rm{sample}}$ = 1000 } & Scaled &   \multirow{2}{1cm}{$N_{\rm{sample}}$ = 1000} & Scaled \\ \\
            & (\%) & (\%) & (\%) & (\%) & (\%) & (\%) \\
		\\
		\hline
		\hline
		\\
		\multirow{2}{2.5cm}[0.5mm]{
 NSBM,\\ $L_{\rm{c,min}}=0.1$ m} & 0.04 & 0.04&  0.3 & 0.4 &  1.5 & 1.7 \\
		 \\ \\
		\multirow{2}{2.5cm}[0.5mm]{ A/M0.05, $N(L_{\rm{c,min}}=0.1$ m)} & 0.2 & 0.3 &  1.2 & 1.4 &  5.3 & 6.1\\
		\\ \\ 
		\multirow{2}{2.5cm}[0.5mm]{ A/M0.04, $N(L_{\rm{c,min}}=0.1$ m)} & 0.3 & 0.3 &  1.5 & 1.7 &  6.6 & 7.7\\
		\\ \\ 
		\multirow{2}{2.5cm}[0.5mm]{ NSBM, \\ $L_{\rm{c,min}}=0.003$ m} & 0.03 & 8.3 & 0.1 & 38.1 &  0.7 & 89.7
		\\
		\\ \\ 
		\multirow{2}{2.8cm}[0.5mm]{ A/M0.05, $N(L_{\rm{c,min}}=0.003$ m)} & 0.2 & 52.6 &  1.2 & 98.0 &  5.3 & 100\\
		\\ \\
		\multirow{2}{2.8cm}[0.5mm]{ A/M0.04, $N(L_{\rm{c,min}}=0.003$ m)} & 0.3 & 60.6 & 1.5 & 99.3 &  6.6 & 100\\
		\\
		 \hline
		 \hline
	\end{tabular}
\end{table}

\subsection{Collision Probability}\label{subsec:colprob}
The probability of one or more collisions between satellites and ASAT fragments over the orbital lifetime of the ASAT debris is shown in Table \ref{tab:results}. As discussed in Section \ref{sec:rebound}, $P_{coll}$ is the cumulative probability that at least one collision will occur between any ASAT debris fragment and any one satellite in our chosen satellite field. The probability is integrated over the lifetime of all fragments, meaning each fragment's contribution to the total probability includes its collision probability up until the fragment de-orbits or the simulation cutoff time is reach (e.g., three years for the India ASAT-like model), whichever occurs first. 
The leftmost column lists each simulation and its corresponding fragmentation model. This is followed by the results for three different satellite density fields, as described in Section \ref{sec:satdis}. For each satellite density field, the collision probability is listed as determined directly from the sample size of 1,000 fragments.  The probability is further scaled to account for the number of fragments predicted by the NSBM above a given size limit, as follows:
\begin{equation}
P_{coll}=1-\exp\left(-\frac{N_{\rm{NSBM}}}{N_{\rm sim}} \Lambda \right)\label{eq:colprob_scaled}
\end{equation}
where $N_{\rm sim}=1000$ for all sims in the table.  
As noted in Section \ref{sec:integrate}, this scaling is possible because the number of fragments in the simulations adequately samples the NSBM distribution functions. Thus, increasing the number of fragments does not change the shape of the distribution functions. As such, each simulated particle's cumulative collision probability scales by a  proportional amount as well.
Recall that the Rayleigh models do not inherently rely on particle size, so the NSBM numbers are also used for their scaling. The similarity between the unscaled and scaled values for cases that use $L_{\rm{c,min}}= 0.1$ m arises because the number of fragments in that model is  close to the 1,000 used in the simulations. There is also a difference in the unscaled probabilities between the two NSBM models. In the realizations studied, the $L_{\rm{c,min}}= 0.003$ m model happens to have a higher fraction of particles that de-orbit right away, which gives rise to the lower unscaled number. Together, the different distributions give insight into the range of outcomes we might expect.

The increase in collision probability from the LEO environment in 2019 to a present-day satcon environment is significant, in some cases as much as two orders of magnitude higher. The sheer number of fragments for sizes down to 3 mm guarantees that a collision of some form will occur, although these may not lead to significant spacecraft damage.
 
To explore this further, we run a higher resolution simulation with 5,000 sampled fragments using the NSBM with 
$L_{\rm{{c,min}}}=3~\rm mm$, for a 65,000-satellite LEO environment. We again derive the fragment numbers $N_{\rm{frag}}$ used for scaling using the NSBM.
We find that the scaled $P_{coll}$ is 88\% for fragments between $L_c =3$ mm and 1 cm in size, 28\% for $L_c>1 \rm~cm$, about 5\% for $L_c > 5 \rm~cm$, and about 2\% for $L_c > 10 \rm~cm$ (these are consistent to a few \% of the values shown in Table \ref{tab:results}).
That is, although the most likely collision to happen is with a small fragment, the probability that a fragment of sufficient size to prompt additional fragmentation events  is non-negligible for ASAT tests similar to those modelled here.

We next calculate the collision probabilities for the aforementioned $L_c$ ranges, but for a higher number of primary collision events. Here, primary refers to only collisions between the ASAT debris fragments and satellites, not collisions involving subsequent fragmentations. Since the collisions in this study follow Poisson statistics, we can extend the probability mass function used for Equations \ref{eq:colprob} and \ref{eq:colprob_scaled} to a non-zero number of events $k$. In other words,
\begin{align}
    P_{coll}(\geq k) &= 1 - \sum_{i=0}^{k-1}\frac{(\Lambda_{\rm{scaled}})^i e^{-\Lambda_{\rm{scaled}}}}{i!}\label{eq:colprob_full}
\end{align}
where $\Lambda_{\rm{scaled}} = \frac{N_{\rm{NSBM}}}{N_{\rm sim}}\Lambda$. We calculate $P_{coll}(\geq k)$ for $k=\{1,3,10\}$ events using the same high-resolution simulation and fragment size ranges explored above. The results are listed in the top panel of Table \ref{tab:highkprob}. We find that $P_{coll}(\geq k)$ drops quickly to zero for $k>1$, and fragments less than a centimetre in size are the most likely to be involved in a collision. The NSBM results thus imply that only a few primary collisions from the ASAT test debris are likely to occur.

We repeat this calculation with a 1000-fragment simulation using the Rayleigh A/M0.04 model, again for a 65,000-satellite LEO environment. We use the same NSBM-derived fragment numbers for scaling as done in the previous calculations. The results are shown in the bottom half of Table \ref{tab:highkprob}. The probabilities are again dominated by fragments smaller than one centimetre in size. As already noted for one or more events (see Table \ref{tab:results} for $k=1$), the Rayleigh model yields much higher collision probabilities than the NSBM.

These results further highlight that, depending on the debris model, the number of  collisions induced by an ASAT test could be of great concern. Exploring again the Rayleigh A/M0.04 model  by scaling the results up to $N_{\rm{frag}}=N([0.003, 1.0])$ (equivalent to $N(L_{\rm{c,min}}=0.003)$), the cumulative collision probabilities for $k$ = 20 and 30 remain non-trivial, with $P_{coll}(\geq k)$ about 75.1\% and 8.5\%, respectively.
\begin{table}[H]
    \centering
	\caption{Scaled probabilities for $\geq k$ primary collisions between ASAT debris and satellites, for $k=\{1,3,10\}$. The top panel uses a high-resolution NSBM simulation with 5,000 debris fragments, $L_{\rm{c,min}}=3~\rm{mm}$, and a full 65,000-satellite density field. The probabilities have been scaled using the NSBM-derived fragments counts for each size range. The bottom panel shows the same calculation but with a 1000-fragment Rayleigh simulation using model A/M0.04. We use the same fragment counts for scaling as the NSBM. The results find a non-negligible chance of multiple primary collisions being induced by the ASAT test, and model A/M0.04 yields higher probabilities than the NSBM overall. Recall that the A/M0.04 model produces fragments that stay in orbit for a longer period of time and thus accumulate higher probabilities.}
	\label{tab:highkprob}
	\begin{tabular}{ccccc}
		\hline
		\hline
		\\
		& \multicolumn{2}{c}{\textbf{NSBM} (scaled)} & &  \\ \\
		$L_c$ range & $P_{coll}(\geq 1)$ & $P_{coll}(\geq 3)$ & $P_{coll}(\geq 10)$ \\
            & (\%) & (\%) & (\%) \\
		\\
		\hline 
		\hline
		\\
        $[0.003,~1.0]$ & 91.5 & 44.7 & 0.02 \\ \\
        $[0.003,~0.01]$ & 88.2 & 35.9 & 0.01 \\ \\
        $[0.01,~1.0]$ & 28.1  & 0.5 &  0.0 \\ \\
        $[0.05,~1.0]$ & 4.7 & 0.0 & 0.0 \\ \\
        $[0.1,~1.0]$ & 2.1 & 0.0 & 0.0 \\
        \\
		\hline
        \hline
		\\
		& \multicolumn{2}{c}{\textbf{Rayleigh} (scaled)} & &  \\ \\
		$N_{\rm{frag}}$ & $P_{coll}(\geq 1)$ & $P_{coll}(\geq 3)$ & $P_{coll}(\geq 10)$ \\
		\\
		\hline 
		\hline
		\\
        $N([0.003,~1.0]$) & 100.0 & 100.0 & 99.9 \\ \\
        $N([0.003,~0.01]$) & 100.0 & 100.0 & 99.5 \\ \\
        $N([0.01,~1.0]$) & 94.1  & 53.6 & 0.07 \\ \\
        $N([0.05,~1.0]$) & 17.1 & 0.09 & 0.0 \\ \\
        $N([0.1,~1.0]$) & 5.2 & 0.0 & 0.0 \\
        \\
		\hline
        \hline
	\end{tabular}
\end{table}

An important caveat is that using $N_{\rm{frag}}$ for scaling the Rayleigh model could overestimate the number of collisions. Indeed, as discussed in Section \ref{subsec:results_deorbits}, the A/M distribution in the NSBM peaks around 0.5 $\rm m^2~kg^{-1}$, much higher than the fixed A/M ratio used in Rayleigh A/M0.04 model. On the other hand, the Rayleigh A/M0.04 simulation shows a better match overall to the Indian ASAT debris reentries. Further work is clearly needed.

Nonetheless, we must keep in mind that the different models explored here cover a wide range of potential outcomes and plausible severity. We also reiterate that each primary collision is able to induce further fragmentation events, even for small fragment sizes; having multiple primary collisions thus increases the risk of additional fragmentations.

\section{Summary and Discussion}\label{sec:conclusions}

In this study, we simulated the evolution of debris produced in kinetic ASAT tests, similar to the 2019 Indian test, and evaluated the probability that such events would lead to collisions with satellites. Different satellite density fields were explored, including different populations of large satellite constellations (satcons). 

Debris fragments were integrated using the \REBOUND\ astrodynamics code to assess the collision probabilities for different fragmentation models. This includes the NASA Standard Breakup Model (NSBM), and a simplified, but instructive Rayleigh velocity distribution model with a constant A/M ratio.

For the NSBM fragmentation models, we found de-orbit timescales that tended to be faster than what was seen with the ASAT test conducted by India in 2019. As such, the fragmentation events that we explore are not only plausible ASAT events, but might underestimate the risks to orbital infrastructure of such behaviour. The Rayleigh models that we used could recreate the de-orbit timescales seen in India 2019 event, depending on the selected fixed A/M ratio.

When using the NSBM, we found that there is $\sim90\%$ probability for fragments with sizes $L_c > 3\rm~mm$ to impact one or more satellites in the 65,000-satellite satcon environment. The 12,000-satellite satcon environment is also high, with a collision probability of about 38\%. These probabilities are both much more significant than that for the 2019 satellite population, which, while non-negligible, is about 8\%. However, even in the case of the 2019 LEO environment, models A/M0.04 and A/M0.05 both show non-negligible collision probabilities above 50\%. Although these impacts do not necessarily mean that they will result in subsequent fragmentation events, they are nonetheless cause for concern. 
 
When focusing on size scales of $L_c > 1~\rm cm$ and $L_c>10~\rm cm$ in the NSBM model, we found that the 65,000-satellite environment still has an approximate 28\% and 2\% chance, respectively, of experiencing one or more collisions, with the latter of particular concern for prompting subsequent large fragmentation events. For the Rayleigh model A/M0.04, we further find that the probability for multiple collisions is non-zero at these size scales. For example, $L_c>1~\rm cm$ yields more than a 50\% chance for three or more collisions occurring over the lifetime of the ASAT debris. However, this probability falls quickly to zero by about ten primary collisions.

With this in mind, the risks associated with non-catastrophic collisions should not be dismissed, as even small fragmentation events contribute to the overall debris population, including non-tracked debris. 

When comparing the two fragmentation methods used here, we see a larger collision probability for the Rayleigh velocity distribution than for the NSBM. The A/M0.04 and A/M0.05 models produce fragments with much longer de-orbit timescale than what occurs when using the NSBM. This just emphasizes the consequences of long-lived debris. 

When interpreting the results, it is important to keep several caveats in mind. First, the satellite density fields are based on plausible satcon configurations, but these are continuously changing. Nonetheless, because ASAT tests will tend to spread fragments over a wide range of altitudes in LEO, the collision rates can be expected to scale with the number of satellites. 

Second, we assumed a fixed cross sectional area for each satellite of $10~\rm m^2$. We note that in reality, the cross sectional area of the satellites depends on the satcon -- a single OneWeb satellite, for example, has dimensions of $1\times1\times3$ metres. We take $10~\rm m^2$ as an average representative value amongst all our satellite systems of interest. The collision rate will again be proportional to the average satellite cross section in each orbital shell. If all shells have comparable satellites, then the expected number of collisions will scale directly with the actual cross section. 

Third, multiple approximations are made in this study. Together with our other assumptions, this includes the extrapolation of atmospheric density at high altitudes, the linear form of Equation \ref{eq:instant_colprob}, and the statistical approximation for the average relative velocity, all of which can be found in Section \ref{sec:rebound}. The simulations in this paper therefore represent only approximations of reality, relying on several simplifications.

Finally, we only focused on one type of kinetic ASAT test, modelled after the 2019 direct-ascent Indian ASAT test (although we explore an additional case in the appendix). We thus did not explore an extensive range of parameter space that could arise from the use of ASAT weapons.

Overall, the results demonstrate that debris-generating ASAT tests are incompatible with providing a safe operating environment in LEO, and are particularly dangerous in a heavily commercialized operating environment that contains satellite constellations. Even events that ensure that fragments de-orbit in less than a year can cause dangerous levels of collisional risk and threaten the continued use and development of LEO by all space users. 

This work was supported in part by the University of British Columbia, the Canada Research Chairs program, a Natural Sciences and Engineering Research Council Discovery grant, and a grant from the Tri-agency New Frontiers in Research Fund. 

\section{Declaration of Competing Interest}
On behalf of all authors, the corresponding author states that there is no conflict of interest.

\begin{appendices}

\section{Russian 2021 DA-ASAT Test}\label{secA1}

On 15 November 2021, the Russian Federation conducted a direct-ascent ASAT weapon test. The target was Cosmos-1408, a 2200 kg defunct Soviet satellite (about three times as massive as that used for the Indian ASAT test), previously used as part of an electronic intelligence system. The impact occurred at an altitude of about 480 km, directly endangering China's Tiangong space station and the International Space Station, forcing crew members to undertake shelter procedures while the ISS passed through the debris cloud. Due to the high altitude of the test where air drag is less influential, the fragment lifetimes will be much longer than those seen in the Indian test. 

Because the Russian test was recent, an analysis of the event is not included in the main text of this article. However, due to the event's impact  on the space environment, we have conducted a preliminary analysis based on the information known at this time, including the distribution of approximately  1500 Cosmos-1408 debris fragments  tracked on \url{space-track.org}. Here, we describe the results of several simulations that we run to assess debris collision probabilities, using the same methods employed for the Indian-like ASAT test.

As done before, satellite density fields are created for the simulations. One density field is based on the satellite catalogue as of November 17th, 2021, for which there are approximately 5700 satellites (active and defunct) in LEO. The second field is the same as the 65,000-satellite satcon field used in the Indian ASAT test simulations.

Table \ref{tab:cosmos1408} shows the orbital parameters for Cosmos-1408 just prior to the impact.  Its high mass along with the assumption of catastrophic fragmentation ensures that a significant amount of debris was created from the test. For example, the NSBM predicts that 1168 pieces of debris with $L_c \geq 10$ cm would be generated for an Indian ASAT-like event. In contrast, the Russian ASAT test generates 2876 of such debris, again based on the NSBM. The potential collision risks posed by this higher number of fragments is compounded by the higher altitude at which the test took place, inducing significantly longer orbital lifetimes for the debris fragments.

\begin{table}[H]
    \centering
	\caption{Orbital parameters of Cosmos-1408, destroyed during the Russia ASAT test.}
	\label{tab:cosmos1408}
	\begin{tabular}{l*2c}
	\\

		\hline
		\hline
		\\
		Altitude of ASAT test & $480$ km \\
        Perigee & 472.5 km  \\
        Apogee & 497.5 km  \\
        Inclination & 82.6°  \\
        Period & 94.3 min  \\
        Semi-major axis & 6862.75 km  \\
        \\
		\hline
        \hline
	\end{tabular}
\end{table}

The Russia ASAT test took place during a solar phase $\phi \approx 1.17 \pi$ (as a reminder, zero and $\pi$ are minima for a full 22-yr cycle, i.e., two solar cycles); this is included in the modelling of the atmospheric gas density and is expected to help in reducing the lifetime of the fragments. 

Simulations are initialized with the NSBM. A kill energy of 150 MJ is assumed, consistent with a catastrophic collision in the NSBM modelling. 
For comparison, we also run a simulation using a Rayleigh velocity distribution and fixed A/M ratio, similar to that done for the Indian ASAT test. We use A/M values of 0.05 m$^2$kg$^{-1}$ and 0.04 m$^2$kg$^{-1}$, which show the best matches between the actual de-orbit times of the tracked Indian ASAT debris and the simulation results. Fragments are integrated until they de-orbit or reach a maximum integration time of 10 years. 
Again, the integrations are done using a satellite density distribution for LEO that reflects the time of the event (see also Section \ref{sec:satdis}). In addition, a second set of simulations are run using a full 65,000-satellite satcon environment, so that the affects of a Russian ASAT-like test on a future LEO environment could be assessed.

\begin{figure}
    \centering
    \includegraphics[scale=0.061, trim={15cm 0cm 15cm 0cm}]{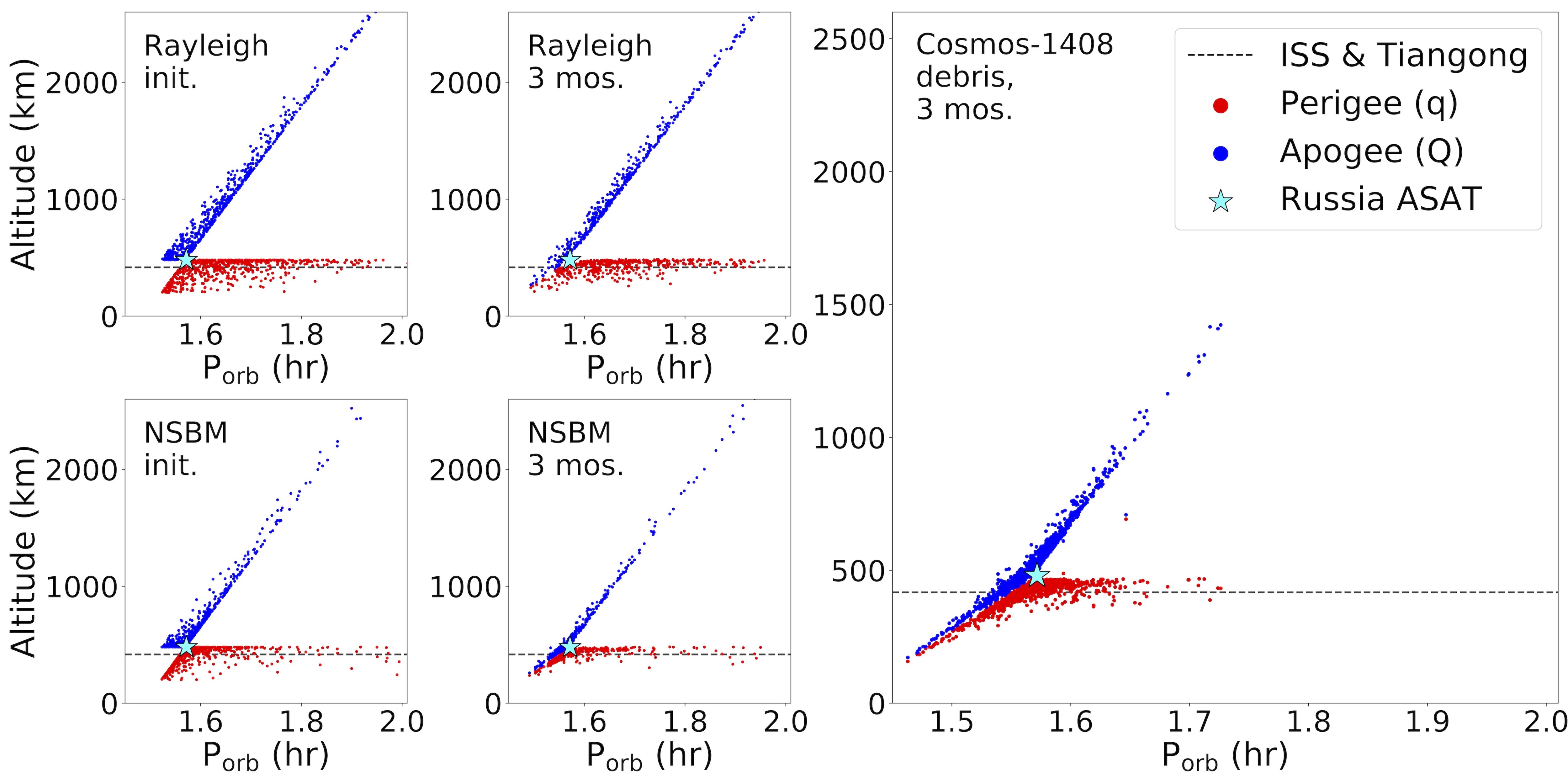}
    \caption{A comparison of distributions between the known fragments from the Russian ASAT test and our implementation of the NSBM and Rayleigh models. The top row from left to right shows Gabbard plots of the initial state of the Rayleigh model and after three months of evolution. The bottom row is similar, but for the NSBM. In addition, the catalogued debris from the Russian ASAT test as of February 10th, 2022, are also shown. A large amount of orbital evolution has already occurred. The altitude and period of the Cosmos-1408 satellite is marked with a blue star. The altitudes of the ISS and Tiangong space station are also indicated. Although the de-orbit timescales of the Rayleigh model is a better fit to the catalogued debris from the Indian ASAT test, the NSBM shows a closer correspondence to the evolution of the Russian ASAT test debris thus far.}
    \label{fig:russia_gabbard}
\end{figure}

\begin{table}
    \centering
	\caption{Collision Probabilities: Russia ASAT-like test}
	\label{tab:resultsrussia}
	\begin{tabular}{l*5c}
	\\
		\hline
		\hline
		\\
		\textbf{Model} & \multicolumn{2}{c}{~~~~\textbf{2021 sats}} & & \multicolumn{2}{c}{~~~~\textbf{65,000 sats}} \\ \\
		   & $P_{coll}$ & $P_{coll}$ & & $P_{coll}$ & $P_{coll}$  \\
		&  $N_{\rm{sample}}$=1000 & Scaled & &  $N_{\rm{sample}}$=1000 & Scaled \\
            & (\%) & (\%) & & (\%) & (\%) \\
		\\
		\hline
		\hline
		\\
		NSBM, $L_{\rm{c,min}}=0.1$ m & 1.7& 4.8& & 6.1 & 16.7\\
		 \\
		A/M0.05, $N(L_{\rm{c,min}}=0.1$ m) & 3.1 & 8.6 & &  11.6 & 29.8\\
		\\
		A/M0.04, $N(L_{\rm{c,min}}=0.1$ m) & 3.6 & 10.1 & &  16.4 & 40.3\\
		\\
		NSBM, $L_{\rm{c,min}}=0.003$ m & 0.7 & 98.9& & 2.0 & 100\\
		\\
		A/M0.05, $N(L_{\rm{c,min}}=0.003$ m) & 3.1 & 100 & & 11.6 & 100\\
		\\
		A/M0.04, $N(L_{\rm{c,min}}=0.003$ m) & 3.6 & 100 & &  16.4 & 100\\ \\
		 \hline
		 \hline
	\end{tabular}
\end{table}

\begin{figure}
    \centering
    \includegraphics[scale=0.043]{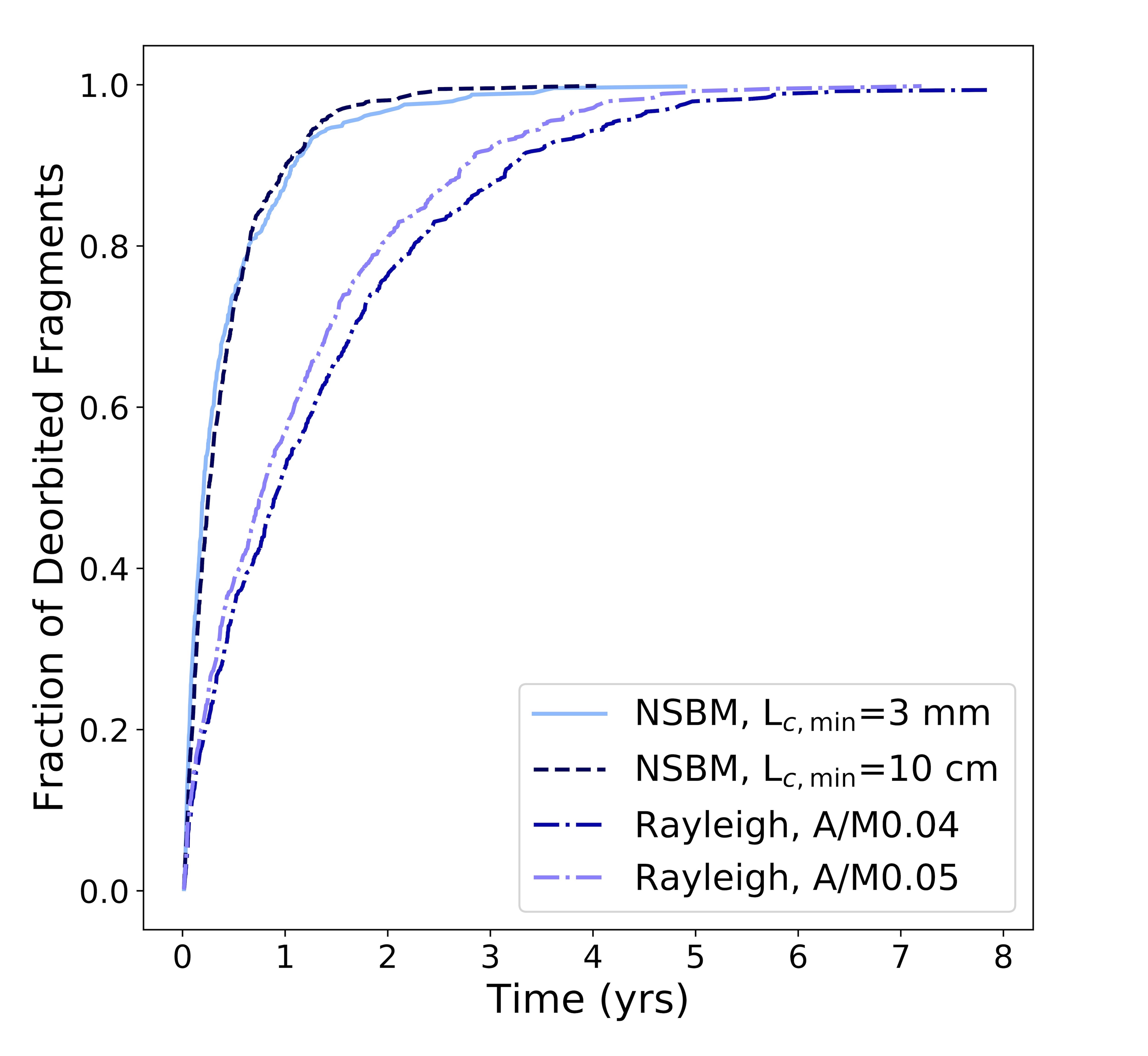}
    \caption{De-orbit timescales for models of the Russian ASAT test debris evolution. Only fragments that last longer than five days are included. Two NSBM simulations are implemented with our chosen minimum characteristic lengths, shown as dark and light blue lines. Rayleigh models A/M0.05 and A/M0.04 are also shown as the dot-dashed royal blue and purple lines. The NSBM leads to short fragment orbital lifetimes compared with the Rayleigh model. The differences in de-orbit outcomes are interpreted as representing the range of possibilities we might reasonably anticipate.}
    \label{fig:russia_deorbit}
\end{figure}

Figure \ref{fig:russia_gabbard} shows Gabbard plots for two different simulations, as well as the actual distribution of the Cosmos-1408 debris. The top row of two panels corresponds to the Rayleigh A/M0.05 model at the initial moment of fragmentation and after three months of integration time, respectively, while the bottom row shows the NSBM model for the initial state and after three months of integration as well. The main right panel is the catalogued Cosmos-1408 debris based on \url{Space-track.org} on February 10th, 2022, about three months since the ASAT test took place. A considerable amount of debris evolution has already occurred, which is most obvious in the morphology at orbital periods less than 1.6 hours. This first three-months of evolution is better captured by the NSBM than the Rayleigh model in this instance. However, we also keep in mind that the Rayleigh model showed higher fidelity to the long-term evolution of the debris for the Indian ASAT test.  There are biases in the catalogued data, and using the NSBM with $L_{\rm{c,min}}=10$ cm does not correct for this behaviour. There  seems to be value in predictions made by both the NSBM and Rayleigh methods, potentially reflecting the initial orbital evolution and the long-term behaviour, respectively.

The range of evolution scenarios for the debris is emphasized in Figure \ref{fig:russia_deorbit}, which shows the de-orbit timescales for simulations of the Russian ASAT test debris using two NSBM initializations and two with the Rayleigh A/M0.05 and A/M0.04 models. The Rayleigh simulations, shown as a dash-dotted purple and royal blue curves, yields a de-orbit timescale that is more consistent with what might be expected based on the Indian ASAT test. Regardless, the figure shows that after roughly a full year of evolution, we should be able distinguish between models.

Table \ref{tab:resultsrussia} lists the unscaled and scaled collision probabilities for the NSBM and Rayleigh modeling of a Russian ASAT-like test in the current satellite environment. We also show probabilities for a satcon environment of 65,000 satellites for comparison. Between the 2021 LEO satellite distribution and the 65,000 satcon environment, there is an increase by a factor of 3-5 in the cumulative collision probability, consistent across all models. As expected, the collision probabilities associated with a Russian ASAT-like test are much higher than for a test similar to the India 2019 event, due to the high altitude of the test and longer de-orbit times. Similar to that seen in our main discussion, the probability of a 10+ cm-sized fragment hitting any one satellite is relatively small, but is still non-negligible at about $10\%$ for a 2021 distribution for A/M0.04; nonetheless, the risk associated with such large collisions should not be dismissed. Of course, if most of these large debris pieces are successfully tracked, then in principle, they can be avoided. However, we again see an effective guarantee that at least one mm-sized fragment will experience a collision with a satellite. Such small fragments are generally non-trackable, so the collision risk is a serious concern. This is alleviated somewhat by such collisions not necessarily disabling a satellite. The number of fragments larger than 3 mm in size is nearly $7\times10^5$ under the NSBM.

\begin{figure}[H]
    \centering
    \includegraphics[scale=0.05]{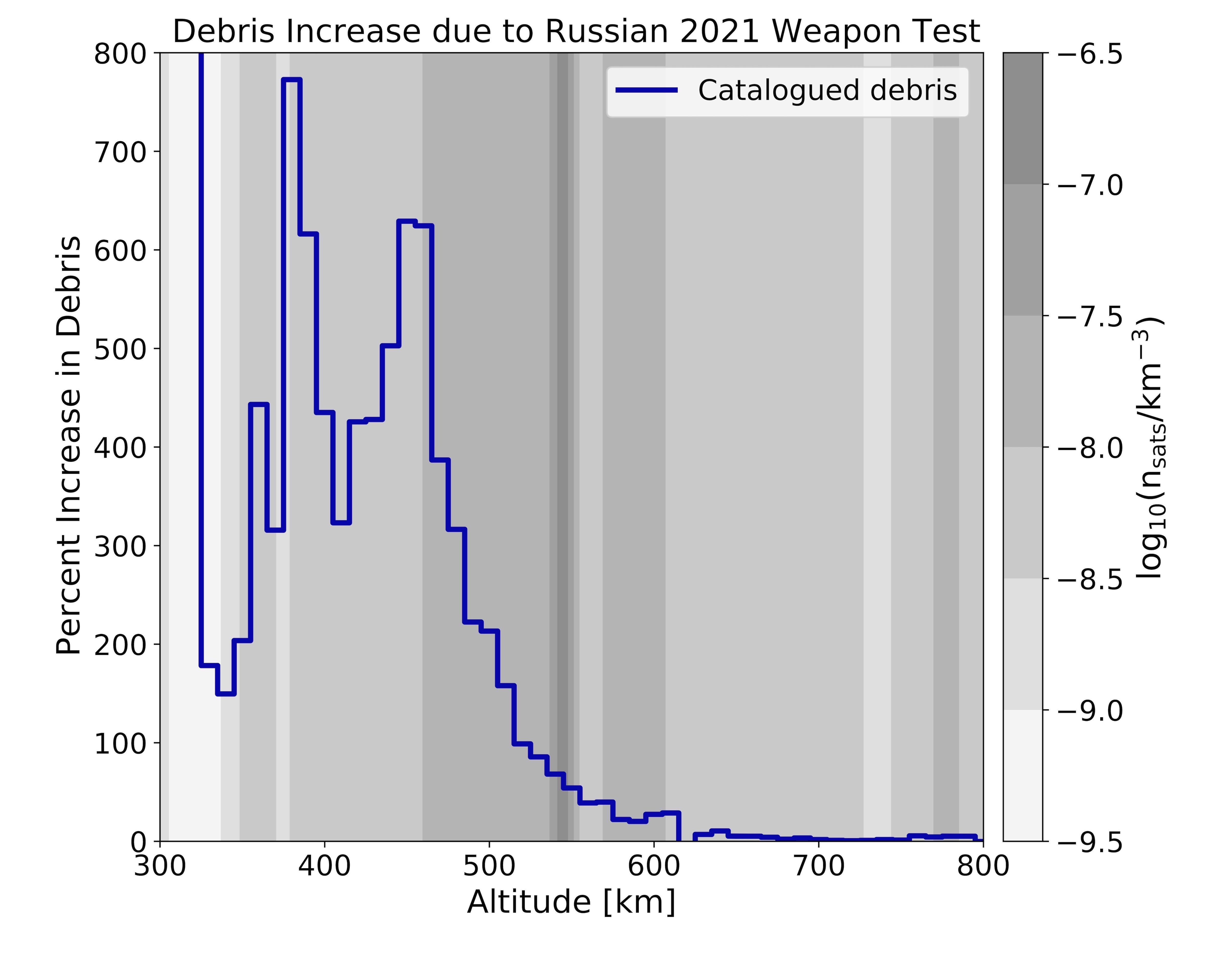}
    \caption{Percent increase of debris density in LEO due to the Russian ASAT weapon test (purple curve). Only fragments that remain in orbit after five days are included. The grey contoured background shows, using a log scale, the number density of satellites. Ten kilometre bins are used for determining the debris densities, calculated using the same method described in section \ref{sec:satdis} for the satellite density field.}
    \label{fig:deb_increase}
\end{figure}

The Russian ASAT test was a large and acute deposit of debris into orbit, with fragments spanning many different LEO altitudes. Figure \ref{fig:deb_increase} shows the percent increase in debris due to the Russian ASAT test. Catalogue data are taken as of January 27th, 2022 and compared with the catalogued data on November 17th, 2021. We specifically filter for debris and ignore other catalogued objects like payloads or rocket bodies. We assume that any significant increases to the debris field over this time span is due solely to the Russian ASAT test. The purple histogram shows the 10 km altitude-binned percentage increase, and the grey shading in the background shows the number density of satellites for each contour, again based on 10 km bins. There is a significant, multi-fold increase in the amount of debris, specifically for altitude bands below 500 km, where there is still a non-negligible satellite number density. This image depicts that despite some beliefs that debris-generating events are only minor contributions to Earth's existing debris field, there is a large LEO impact associated with acute events like ASAT tests.

\end{appendices}

\end{document}